\documentclass[twocolumn]{aastex631}

\accepted{by The Astrophysical Journal on 22 January 2025}

\usepackage{graphicx}	
\usepackage{amsmath}	
\usepackage{textcomp,gensymb}	
\usepackage{overpic}

\graphicspath{ {./images/} }

\newcommand{\isois}{IS$\sun$IS}

\newcommand{\app}{$\sim$}
\newcommand{\gt}{\textgreater}

\newcommand{\kms}{$kms^{-1}$}
\newcommand{\Rs}{$ R_{\odot}$}
\newcommand{\tf}{\footnotesize}

\begin{document}
\begin{sloppypar}

\title{Radial dependence of ion fluences in the 17 July 2023 SEP event \\ from Parker Solar Probe to STEREO and ACE}

\correspondingauthor{Gabriel D. Muro}
\email{gmuro@caltech.edu}

\author[0000-0003-0581-1278]{G.D. Muro}
\affiliation{California Institute of Technology, Pasadena, CA 91125, USA}


\author[0000-0002-0978-8127]{C.M.S. Cohen}
\affiliation{California Institute of Technology, Pasadena, CA 91125, USA}

\author[0000-0002-9246-996X]{Z. Xu}
\affiliation{California Institute of Technology, Pasadena, CA 91125, USA}

\author[0000-0002-0156-2414]{R.A. Leske}
\affiliation{California Institute of Technology, Pasadena, CA 91125, USA}


\author[0000-0003-2134-3937]{E.R. Christian}
\affiliation{NASA/Goddard Space Flight Center, Greenbelt, MD 20771, USA}

\author[0000-0002-3840-7696]{A.C. Cummings}
\affiliation{California Institute of Technology, Pasadena, CA 91125, USA}

\author[0000-0002-3677-074X]{G. De Nolfo}
\affiliation{NASA/Goddard Space Flight Center, Greenbelt, MD 20771, USA}

\author[0000-0002-7318-6008]{M. I. Desai}
\affiliation{Southwest Research Institute, San Antonio, TX 78228, USA}
\affiliation{University of Texas, San Antonio, TX 78249, USA}

\author[0000-0002-5456-4771]{F. Fraschetti}
\affiliation{University of Arizona, Tucson, AZ 85721, USA}
\affiliation{Center for Astrophysics $|$ Harvard $\&$ Smithsonian, Cambridge, MA 02138, USA}

\author[0000-0002-0850-4233]{J. Giacalone}
\affiliation{University of Arizona, Tucson, AZ 85721, USA}

\author[0000-0001-9178-5349]{A. Labrador}
\affiliation{California Institute of Technology, Pasadena, CA 91125, USA}

\author[0000-0001-6160-1158]{D.J. McComas}
\affiliation{Department of Astrophysical Sciences, Princeton University, Princeton, NJ 08544, USA}

\author[0000-0003-4501-5452]{J.G. Mitchell}
\affiliation{NASA/Goddard Space Flight Center, Greenbelt, MD 20771, USA}

\author[0000-0003-1960-2119]{D.G. Mitchell}
\affiliation{Johns Hopkins University Applied Physics Laboratory, Laurel, MD 20723, USA}

\author[0000-0002-8111-1444]{J. Rankin}
\affiliation{Department of Astrophysical Sciences, Princeton University, Princeton, NJ 08544, USA}

\author[0000-0002-3737-9283]{N.A. Schwadron}
\affiliation{University of New Hampshire, Durham, NH 03824, USA}

\author[0000-0002-3093-458X]{M. Shen}
\affiliation{Department of Astrophysical Sciences, Princeton University, Princeton, NJ 08544, USA}

\author[0000-0002-2825-3128]{M.E. Wiedenbeck}
\affiliation{NASA/Jet Propulsion Laboratory, Pasadena, CA 91011, USA}

\author[0000-0002-1989-3596]{S.D. Bale}
\affiliation{University of California Berkeley, Berkeley, CA 94720, USA}

\author[0000-0002-4559-2199]{O. Romeo}
\affiliation{University of California Berkeley, Berkeley, CA 94720, USA}

\author[0000-0002-8164-5948]{A. Vourlidas}
\affiliation{Johns Hopkins University Applied Physics Laboratory, Laurel, MD 20723, USA}




\begin{abstract}

In the latter moments of 17 July 2023, the solar  active region 13363, near the southwestern face of the Sun, was undergoing considerable evolution, which resulted in a significant solar energetic particle (SEP) event measured by Parker Solar Probe’s Integrated Science Investigation of the Sun (\isois) and near-Earth spacecraft. Remote observations from GOES and CHASE captured two M5.0+ solar flares that peaked at 23:34 and 00:06 UT from the source region. In tandem, STEREO COR2 first recorded a small, narrow coronal mass ejection (CME) emerging at 22:54 UT and then saw a major halo CME emerge at 23:43 UT with a bright, rapidly expanding core and CME-driven magnetic shock with an estimated speed of \app1400 \kms. Parker Solar Probe was positioned at 0.65 au, near-perfectly on the nominal Parker spiral magnetic field line which connected Earth and the active region for a 537 \kms ambient solar wind speed at L1. This fortuitous alignment provided the opportunity to examine how the SEP velocity dispersion, energy spectra, elemental composition, and fluence varied from 0.65 to 1 au along a shared magnetic connection to the Sun. We find a strong radial gradient, which is best characterized for H and He as $r^{-4.0}$ and most surprisingly is stronger for O and Fe which is better described by $r^{-5.7}$.

\end{abstract}




\section{Introduction} \label{intro}

Solar energetic particle (SEP) events are periods of elevated intensities of high-energy particles typically associated with solar flares, coronal mass ejections (CME), stream interaction regions, corotating interaction regions, and other energetic solar events \citep{cane1986,reames1999,kallenrode2003}. These particles can pose a hazard to astronauts and spacecraft \citep{sihver2021} along with causing economic damage on Earth \citep{eastwood2018}. Characterizing SEP events is crucial for understanding space weather and its impact on space missions and satellite operations, in addition to a broader understanding of the origin of other high kinetic energy particles such as cosmic rays.

The difficulty of connecting near-Sun activity to SEP events measured in-situ is that the SEPs exhibit no obvious characteristics that can be detected via remote sensing methods, especially at planetary distances. The unfortunate caveat of in-situ measurements is that we are often limited to a single observation point to work from, so the rare opportunities to measure SEPs at different points within the solar system are important for sifting out the key transport mechanisms that affect a given parcel of particles. Notable previous multi-spacecraft events that examined radial effects included \cite{lario2006} through the IMP 8 and HELIOS era, 4 June 2011 and 17 Nov 2011 \cite{lario2013}, 17 April 2021 \citep{dresing2023}, and 14 March 2022 \citep{walker2025}. More rare are SEP events which are observed by spacecraft with small longitudinal separation and radially separated by interplanetary distances, such as 28 April 1978 and 21 June 1981 \cite{lario2006} as well as the 17 July 2023 event described in this study. 

The major benefit of this 17 July 2023 SEP event is that it occurred during the era of spacecraft particle detectors with the ability to examine compositional variations in heavy ions, whereas earlier instruments were limited to protons, electrons and He. So the study of radial evolution of SEPs in the era of Parker Solar Probe (PSP) is key to determining the ion dependent favorable interplanetary conditions for transport along the Parker spiral magnetic field and the ability to test diffusive models for particle transport.

\section{Spacecraft configuration and solar activity}

\begin{figure}[htbp]
    \centering
    \includegraphics[width=1.0\linewidth]{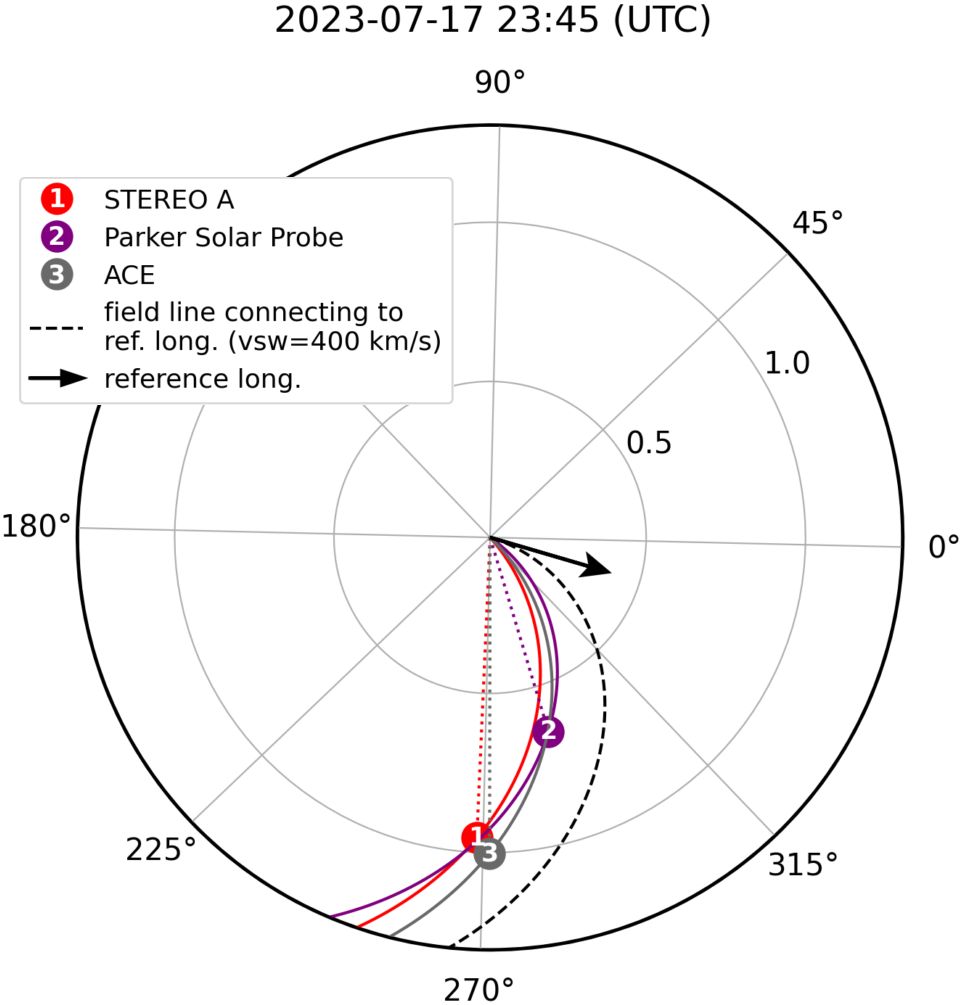}
    \caption{Spacecraft configuration of (1) STEREO A in red, (2) PSP in violet, and (3) ACE in grey prior to the onset of the SEP event from the Solar-MACH model \citep{gieseler2023}. The black reference arrow designates the footpoint of the flares and field line projected to 1 au. The nominal Parker spiral field lines based on the measured solar wind speed are plotted and colored to match their associated spacecraft. See Table \ref{tab-solarmach} for specifications.}
    \label{f-solarmach}
\end{figure}

\begin{table}
\begin{center}
\caption{Parameters of spacecraft configuration in Figure \ref{f-solarmach}}
\label{tab-solarmach}
\hspace*{-1.5cm}\begin{tabular}{|l|c|c|c|}
  \hline
  \tf \textbf{} & \tf \textbf{STA} & \tf \textbf{ACE} & \tf \textbf{PSP} \\
  \hline
  \tf Carrington longitude [\degree] & \tf 268.8 & \tf 271.2 & \tf 288.0 \\
  \hline
  \tf Carrington latitude [\degree] & \tf 4.4 & \tf 4.5 & \tf 3.2 \\
  \hline
  \tf Heliocentric distance [au] & \tf 0.96 & \tf 1.01 & \tf 0.65 \\
  \hline
  \tf Solar wind speed [km/s] & \tf 538 & \tf 538 & \tf 417 \\
  \hline
  \tf Magnetic footpoint Carr. long. & \tf 313.1 & \tf 317.8 & \tf 326.6 \\
  \hline
\end{tabular}
\end{center}
\end{table}

On 17-18 July 2023, the largest SEP event in the 16th orbit of PSP occurred when the spacecraft was nearly aligned with the Solar Terrestrial Relations Observatory (STEREO), and Advanced Composition Explorer (ACE) along a common Parker spiral.  This alignment allows for the measurement of particles from the same SEP event as they travel at approximately 10\% of the speed of light, passing each spacecraft from 0.65 to \app 1.0 au. The ability to analyze particles from the same SEP event at multiple spacecraft locations provides an unprecedented opportunity to study the evolution of SEP event characteristics as it propagates radially, nearly indepdendent of longitudinal effects, through interplanetary space. About 29 hours after the initial burst of SEPs, the interplanetary shock driven by the coronal mass ejection (CME) associated with the event reached PSP and increases in lower energy ions were measured.

The uniqueness of the spacecraft configuration is shown in Figure \ref{f-solarmach}, which displays the location of all 3 spacecraft in a 2-D plane and the nominal Parker spirals based on measured solar wind velocities.  Table \ref{tab-solarmach} lists the related parameters as well as the negligible latitudinal separation of 1.3\degree. The relatively minor magnetic footpoint separation of \app11\degree\ longitude indicates that all three spacecraft are magnetically connected to the same general region on the solar surface, providing confidence that the measured SEPs are from the same event.

\begin{figure}[htbp]
    \centering
    \includegraphics[width=1.0\linewidth]{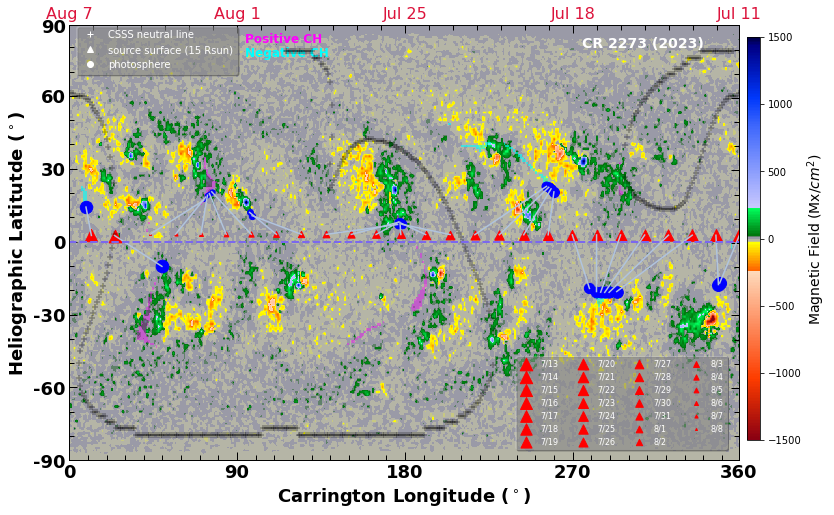}
    \caption{PFSS model derived magnetic field map from SDO/HMI \citep{poduval2014}. Red triangles mark the magnetic footpoint of PSP at 15 Rs, blue circles indicate the coresponding PFSS-determined footpoint location in the photosphere, and the black crosses outline the heliospheric current sheet boundary.}
    \label{f-pfss}
\end{figure}

Figure \ref{f-pfss} shows the SDO/HMI photospheric magnetic field map for the relevant Carrington rotation 2273, the magnetic footpoints of PSP at 15 \Rs, and the corresponding photospheric footpoints calculated from the  potential field source surface (PFSS) model\footnote{Available at: \url{https://spp-isois.sr.unh.edu/psp_footpoints/Plots/}}. The solar activity emerged from active region (AR) 13363, located on this map at (-22\degree, 285\degree)\footnote{Heliographic latitude, Carrington longitude} and PSP was clearly well-connected to it for several days.  Also shown is the calculated position of the heliospheric current sheet (HCS).  The complex shape of the HCS, which crosses the solar equator at roughly 210\degree longitude, had persisted in this general configuration for nearly 6 months. Although, the bulge-like shape in the HCS at (+15\degree, 315\degree) seems to have only begun to form during the previous Carrington rotation, suggesting some more localized evolution during the observational period of the SEP event.

\begin{figure}[htbp]
    \centering
    \begin{interactive}{animation}{movie_COR_EUVI.v2.mp4}
    \includegraphics[width=0.9\linewidth]{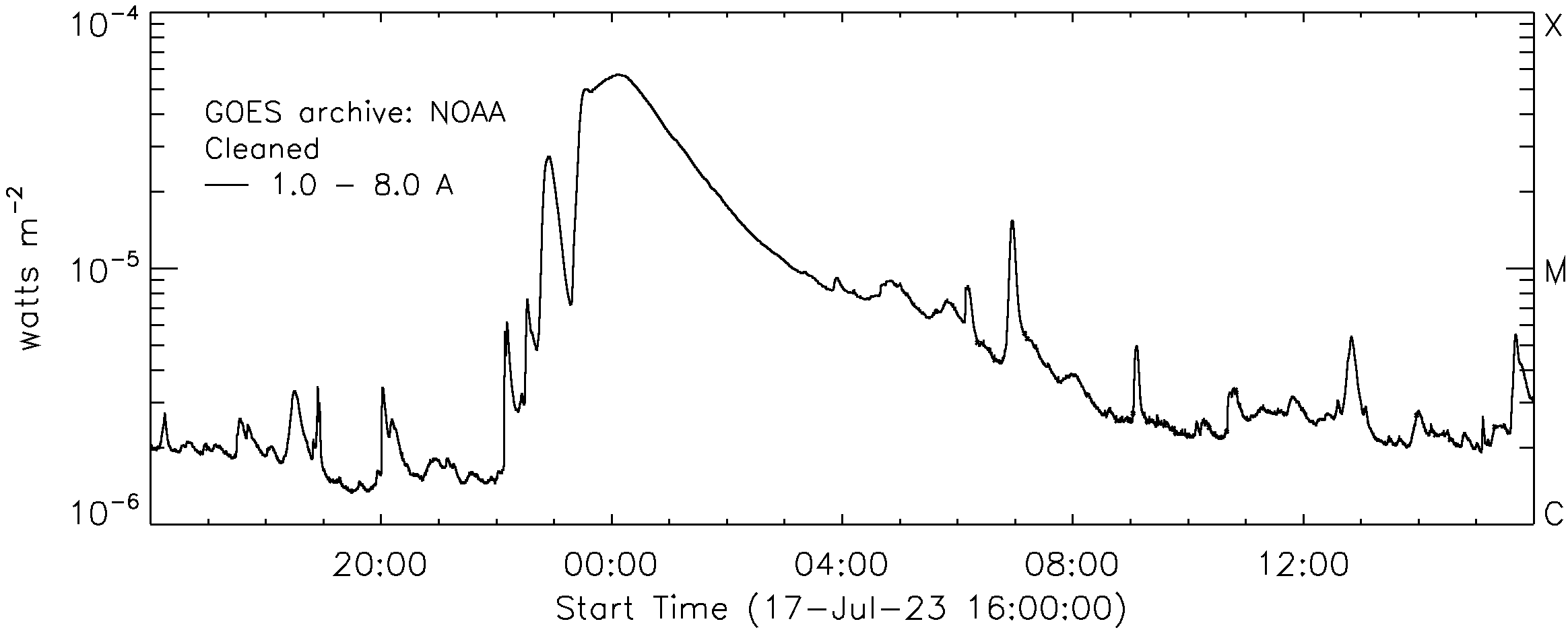}
    \includegraphics[width=0.9\linewidth]{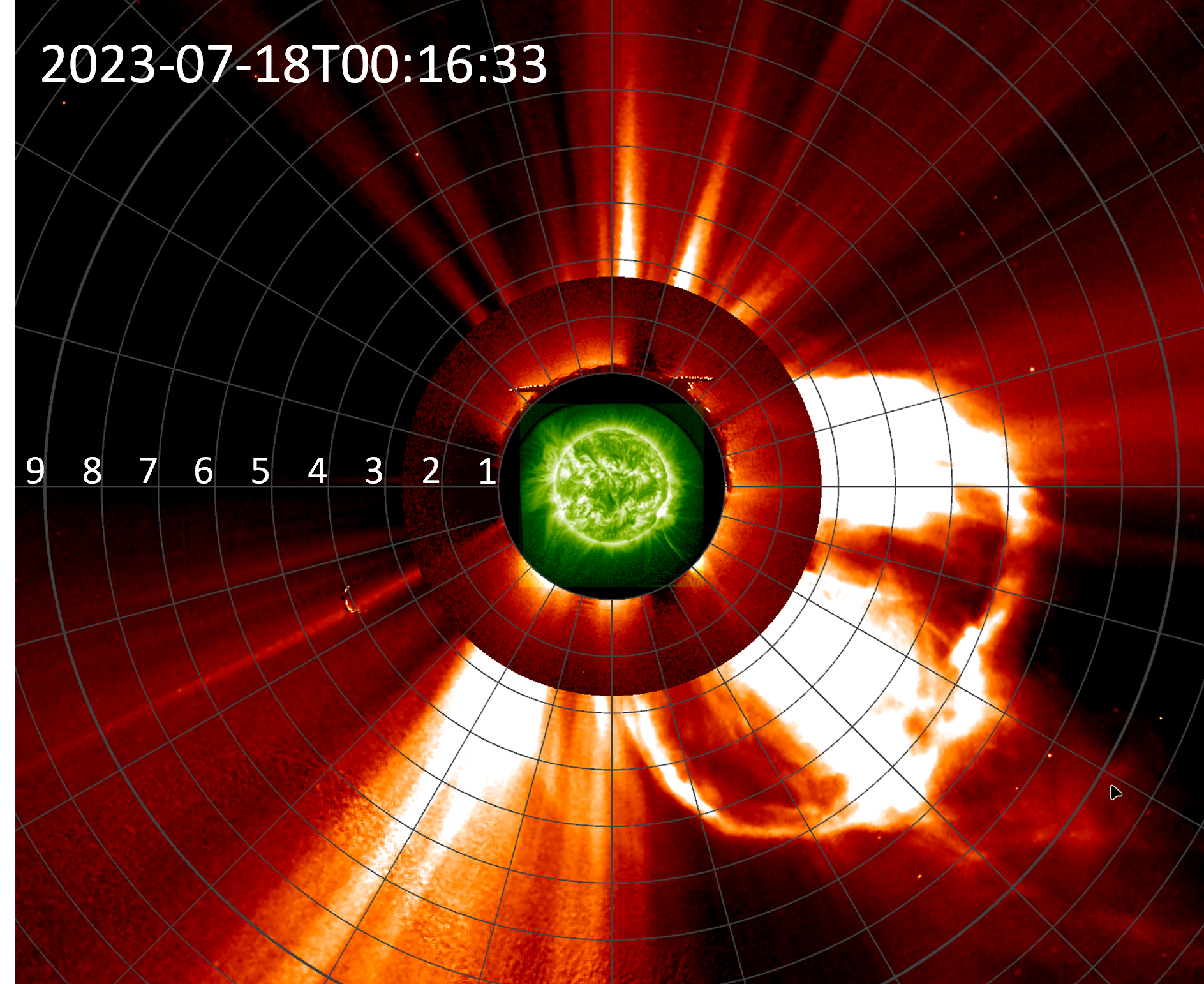}
    \end{interactive}
    \caption{(Top) Time evolution from GOES-16 spacecraft of 1 to 8 \r{A} for 1 second X-ray flux (left y-axis) and flare category according to GOES classification (right y-axis). (Bottom) Composite image of STEREO COR2 and COR1, colored red, shows white light coronagraph images from 1 to 9 \Rs. EUVI 195 ultraviolet light image of the solar disk, colored green, is also shown for context. Note: The supplemental animation begins on 17 July 2023 at 22:00 and continues over the next 10 hours to show two CMEs in quick succession erupting from AR 13363.}
    \label{f-x-ray_cme}
\end{figure}

AR 13363 exhibited notable activity preceding the SEP event. It started with an M5.0 flare that began at 22:54 UT on 17 July 2023 based on GOES-16 X-ray measurements in Figure \ref{f-x-ray_cme}, followed by a small, narrow CME. Then, at 23:17 UT a larger M5.6 flare occurred, with an extended period of X-ray emission for several hours, and a large halo CME erupted from the southwestern area as shown by STEREO COR2's white-light coronagraph in Figure \ref{f-x-ray_cme}.

\section{In-situ energetic particle instrumentation}

This study utilizes the energetic particle instruments on PSP, STEREO, and ACE.  While there are differences in the temporal, energy and mass resolution of the instruments, they generally measure similar energy ranges between 1 and tens of MeV/nuc.  The sensors are capable of measuring energetic ions from H to Fe (with the exception that ACE/SIS does not measure protons) and here we concentrate on H and He (referred to as ``light ions") and O and Fe (``heavy ions").  The details of the instrumentation on each spacecraft is described below.

\subsection{\isois\ onboard PSP}

PSP is a flagship NASA mission designed to study the upper corona and inner heliosphere in-situ \citep{fox2016} and the \isois\ \citep{mccomas2016} instrument suite is designed to study energetic particles at close proximity to the Sun after its launch in 2018. \isois\ is comprised of a low energy instrument, EPI-Lo \citep{hill2017}, which uses the time-of-flight verses residual energy technique to measure \app0.020 to 10 MeV ions and contains 80 apertures spread across a hemisphere, and a high energy instrument, EPI-Hi \citep{wiedenbeck2017}, which covers roughly 1 to \gt100 MeV/nucleon, depending on chemical species.  EPI-Hi is further partitioned into the Low-Energy Telescope (PSP/LET) and High-Energy Telescope (PSP/HET), both of which utilize the standard dE/dx verses residual energy technique to measure ions in an approximate energy range of 1 to 20 MeV/nucleon for PSP/LET and 10 to 100 MeV/nucleon for PSP/HET.

PSP/LET has three apertures, PSP/LET A is oriented 45\degree\ away from the spacecraft-Sun line in the approximate sunward direction of a Parker spiral magnetic field line and, PSP/LET B, points 180\degree\ away from PSP/LET A.  The third aperature, PSP/LET C, points 90\degree\ between the other two along the ecliptic plane. PSP/HET has two apertures, PSP/HET A is oriented 20\degree away from the spacecraft-Sun line and PSP/HET B is pointed 180\degree\ from the PSP/HET A direction. This study primarily utilizes the sunward pointing apertures (PSP/LET A and PSP/HET A).

\subsection{Low and High Energy Telescopes onboard STEREO}

STEREO \citep{kaiser2007} formerly consisted of two NASA spacecraft, Ahead and Behind, launched in 2006 to provide stereoscopic observations of the Sun. Sadly, STEREO Behind suffered hardware damage and was lost in 2014, prior to the launch of PSP, so all STEREO references in this study are for the surviving STEREO Ahead (STA) spacecraft. STEREO's instrument suite includes the Sun Earth Connection Coronal and Heliospheric Investigation imaging suite of remote ultraviolet (EUVI) and white-light (COR1, COR2) coronagraph cameras which provide the data shown in Figure \ref{f-x-ray_cme}. The particle detectors onboard STEREO used in this study are the Low Energy Telescope (STA/LET) \citep{mewaldt2007} and High Energy Telescope (STA/HET) \citep{vonRosenvinge2008} as part of the In-situ Measurements of Particles and CME Transients (IMPACT) suite of instruments. 

STA/HET is composed of 9 solid-state detectors to measure SEPs in the energy range of 13 to 100 MeV/nucleon in \app 11 energy intervals and arranged in stacks oriented in the sunward direction with 55\degree\ fields of view. STA/LET is composed of 14 solid-state detectors to measure SEPs in the energy range of 3 to 30 MeV/nucleon in \app 12 energy intervals and arranged in two fans directed in the sunward and anti-sunward directions with wide fields of view 123\degree\ in the ecliptic plane by 29\degree out of the ecliptic plane. STA/HET's ion measurements are limited to H, while STA/LET has observations of H, He, O and Fe.

\subsection{Solar Isotope Spectrometer (ACE/SIS) onboard ACE}

ACE \citep{stone1998} is a NASA spacecraft launched in 1997 to study particle composition, space weather, and the solar wind at the semi-stable L1 Lagrange point. ACE carries Solar Isotope Spectrometer (ACE/SIS) particle detector instrument to measure the energy spectra and elemental composition of SEPs to give insight into their origin and acceleration mechanisms.

ACE/SIS is optimized for SEPs with energies ranging from \app10 to 100 MeV/nucleon via solid-state silicon detectors that are able to identify energetic nuclei from He to Ni over this energy range. The time cadence of ACE/SIS is more limited in comparison to PSP/HET, PSP/LET, and STA/LET with intensities available every 256 seconds (versus every 60 seconds for PSP and STA) and is not designed to detect anisotropies.  During this SEP event, ACE and STEREO are quite close to each other, thus their measurements provide consistency checks as well as ACE/SIS is able to extend the He, O, and Fe measurements to higher energies than is possible from STA/LET alone.  Additionally, STA/LET and STA/HET provide the H measurements that ACE/SIS does not have.

\section{Particle measurements and analysis}

\subsection{Velocity dispersion}

\begin{figure}[htbp]
    \centering
    \begin{overpic}[width=1.00\linewidth]{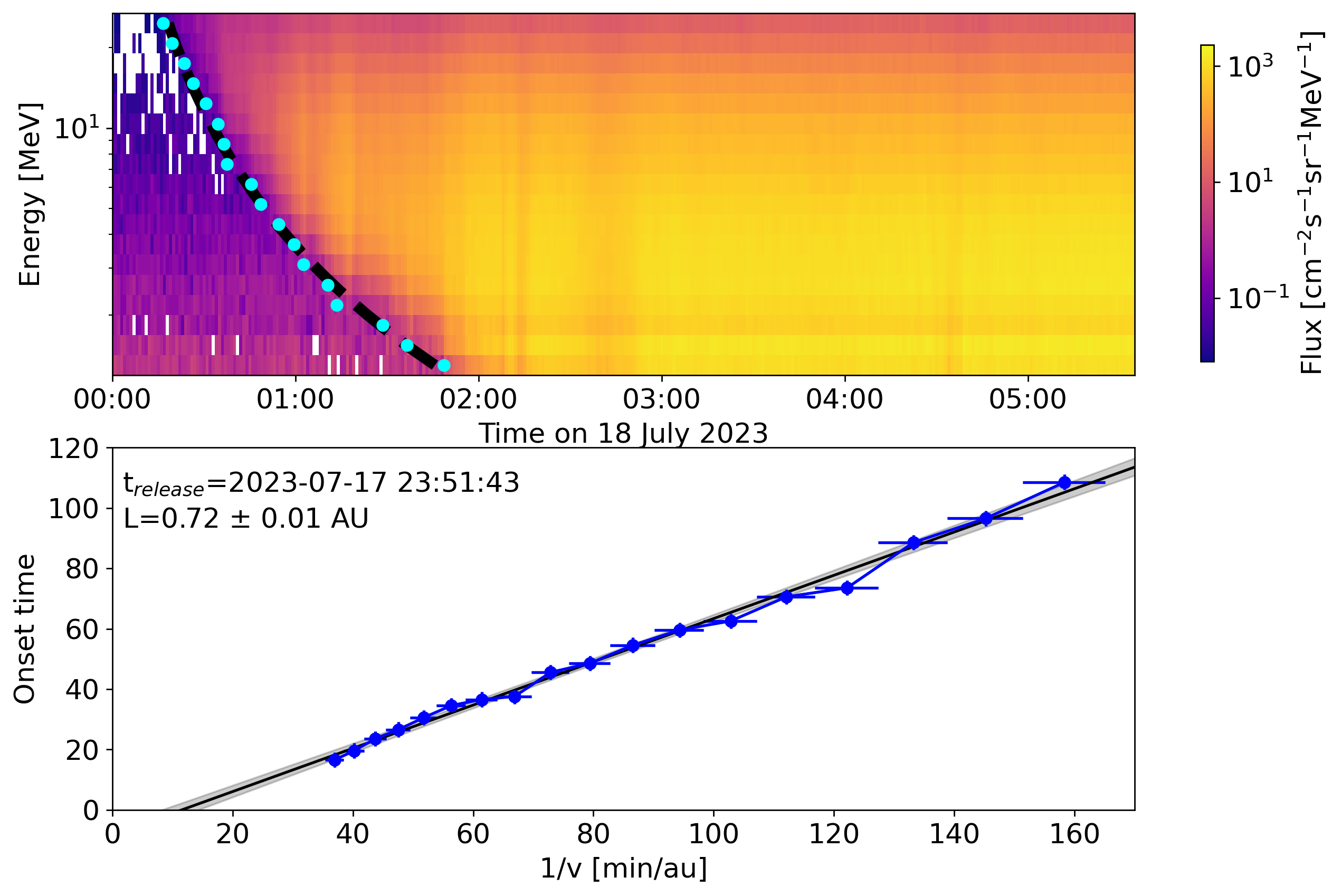}
        \put(65,42){\color{black}\small PSP/HET}
        \put(65,10){\color{black}\small PSP/HET}
    \end{overpic}
    \begin{overpic}[width=1.00\linewidth]{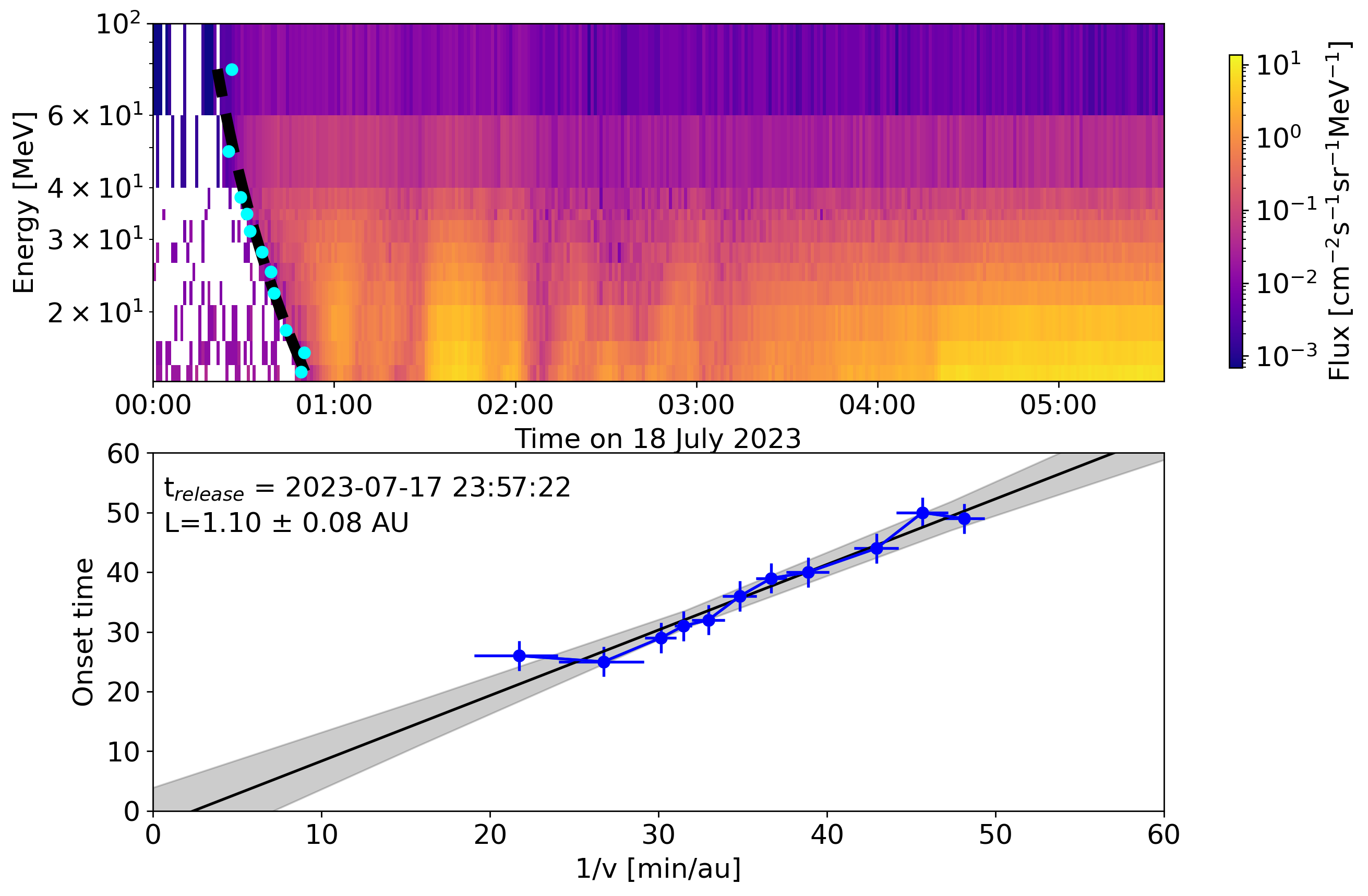}
        \put(65,42){\color{black}\small STA/HET}
        \put(65,10){\color{black}\small STA/HET}
    \end{overpic}
    \caption{(Top row) Proton intensity spectrograms measured at 1 minute intervals by the (left) PSP/HET and (right) STA/HET instruments for the first 6 hours of 18 July 2023. The cyan dots are automated fit points for the arrival of protons at a given energy. (Bottom row) Velocity dispersion analysis of the spectrogram fit points, providing the particle solar release time (t0) and the pathlength (L).}
    \label{f-vda_psp}
\end{figure}

In order to estimate the release time of the SEPs, we performed velocity dispersion analysis (VDA) by automatically selecting sharp increases in intensity using the method of \cite{xu2020}. VDA assumes that SEPs of all energies were injected at the same time and that the first particles measured are those which traveled a nearly scatter-free path with negligible cross-field diffusion to reach the spacecraft. Figure \ref{f-vda_psp} shows the result from the PSP/HET A proton data, where the onset time as a function of inverse velocity results in a path length of 0.72 au, which is consistent with a Parker spiral-like path that would be expected compared to the radial PSP distance of 0.65 au, and a release time of 23:51 UT. Also shown in Figure \ref{f-vda_psp} is the VDA result for STA/HET A proton data, where the release time is estimated as 23:57 UT and the overall path length is 1.10 au compared to the radial STA distance of 0.96 au, the increasing uncertainties is expected as increased scattering occurs over interplanetary distances.

It is important to keep in mind that VDA should be treated only as a guide, since simultaneous injection, comparable spatial transport and acceleration time scale for a single location is not ensured for even the first arriving particles. Thus, the differences in SEP path lengths and release times at PSP and STA are not a concern, especially when considering the larger uncertainties for release time for STA/HET as shown in Figure \ref{f-vda_psp}. The larger uncertainty suggests that particle scattering during the last \app0.35 au of the SEP path have affected the accuracy of VDA. More importantly, the estimated release times from both spacecraft's observations indicate that the release time of the SEP event occurred slightly over 40 minutes after the second M-class flare occurred when the halo CME was at 6+ \Rs.

\subsection{Fluence spectra}

\begin{figure}[htbp]
    \centering
    \includegraphics[width=1.0\linewidth]{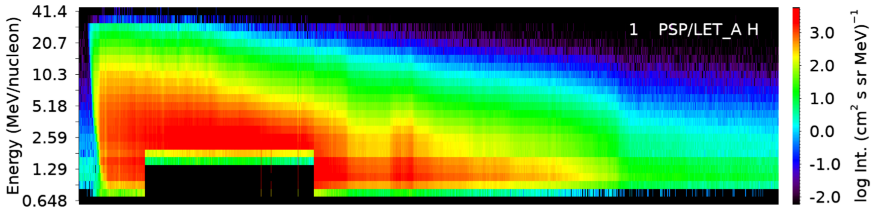}
    \includegraphics[width=1.0\linewidth]{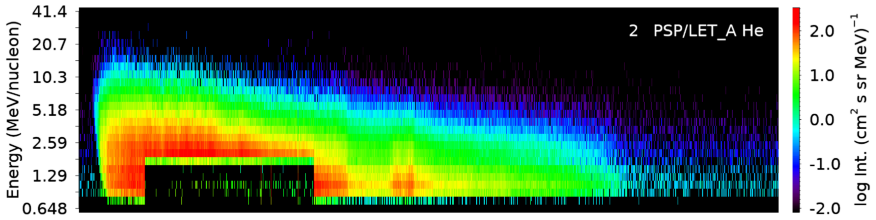}
    \includegraphics[width=1.0\linewidth]{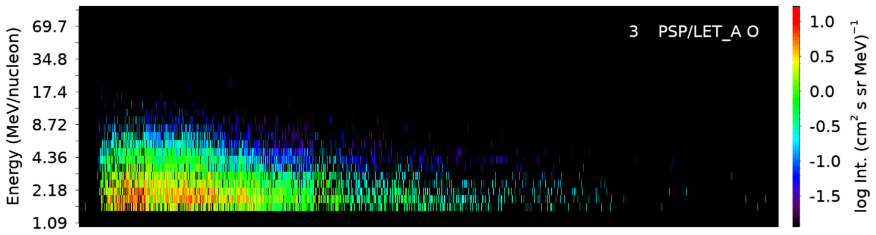}
    \includegraphics[width=1.0\linewidth]{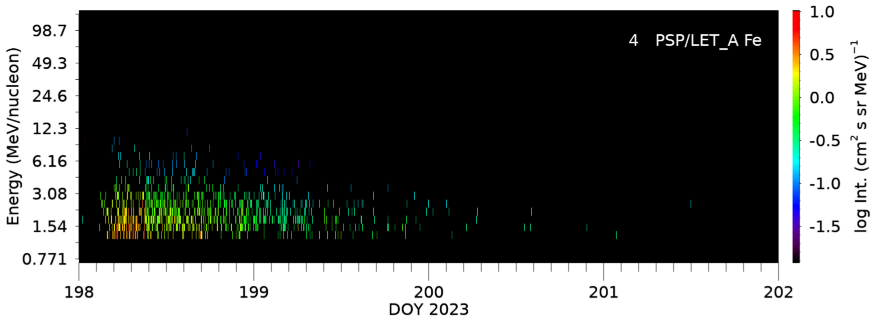}
    \caption{Energy spectrograms measured by PSP/LET A for: (1) protons, (2) helium, (3) oxygen, and (4) iron. The cut out in the H and He intensities at lower energies from 18 July 2023 at 08:03 to 19 July 2023 at 07:15 is due the instrument being in dynamic threshold mode 1 \citep{cohen2021}.}
    \label{f-psp_let_spectra}
\end{figure}

\begin{figure}[htbp]
    \centering
    \includegraphics[width=1.0\linewidth]{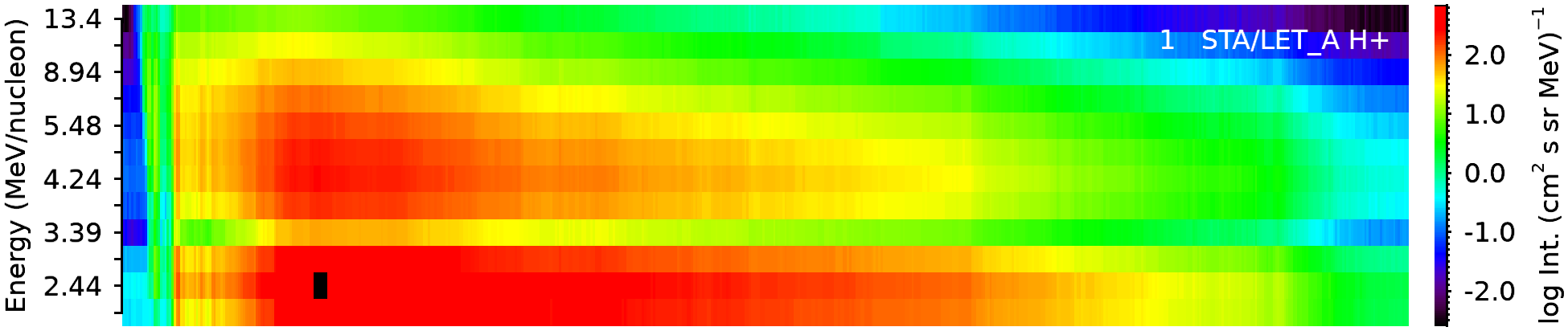}
    \includegraphics[width=1.0\linewidth]{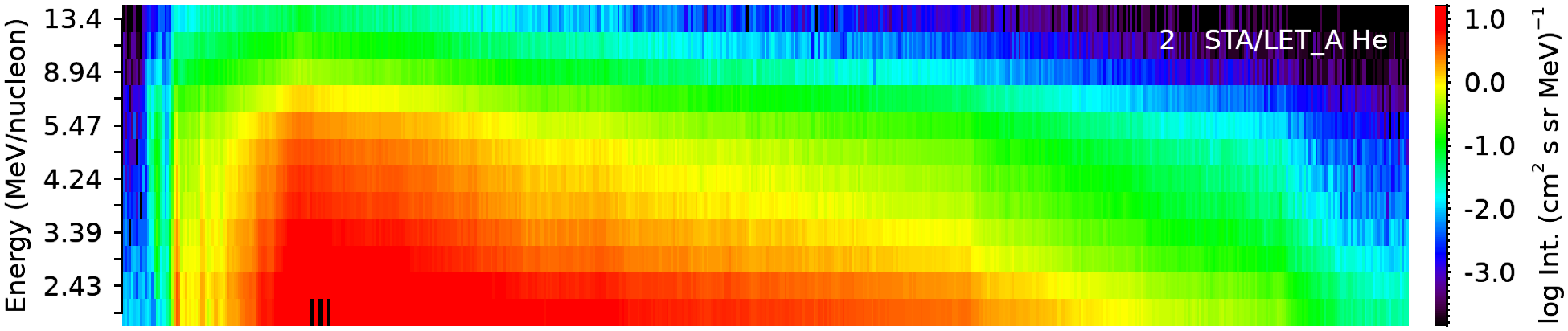}
    \includegraphics[width=1.0\linewidth]{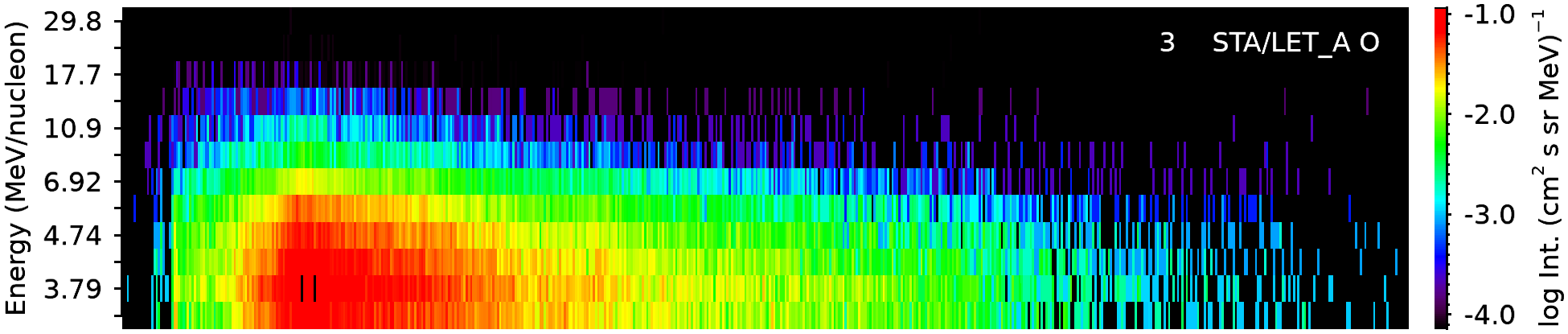}
    \includegraphics[width=1.0\linewidth]{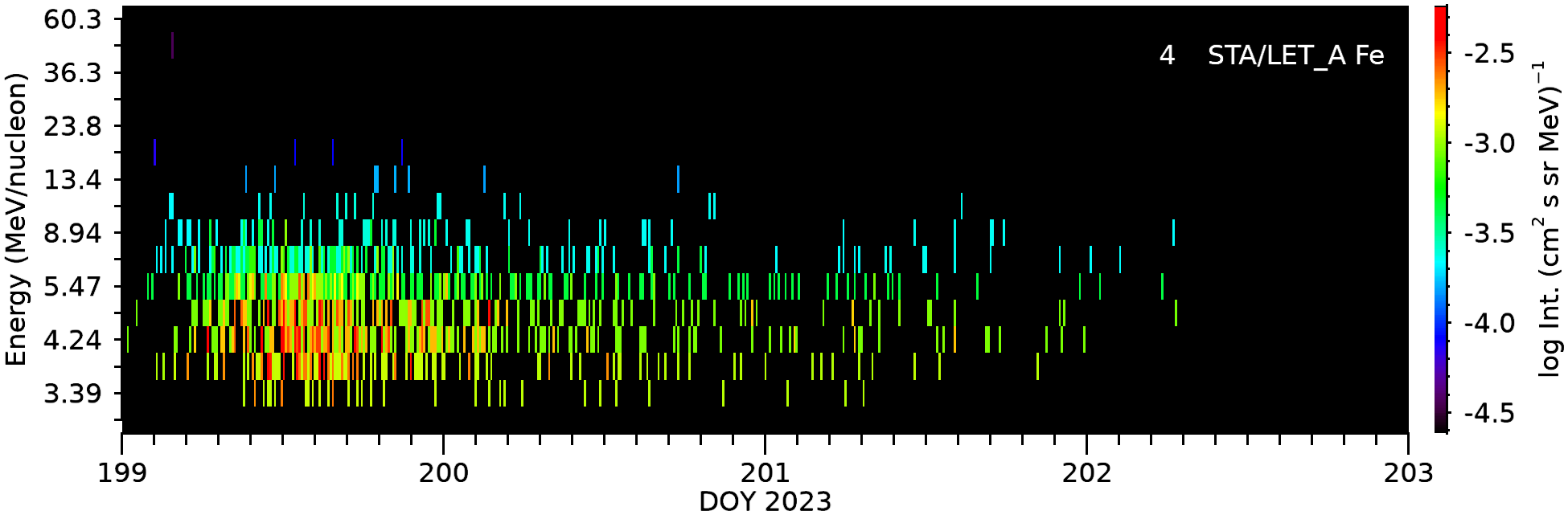}
    \caption{Energy spectrograms measured by STA/LET A for: (1) protons, (2) helium, (3) oxygen, and (4) iron at 10 minute integrations. Note reduced proton flux at \app3.38 is a known instrumental effect.}
    \label{fig:stereo_let}
\end{figure}

\begin{figure}[htbp]
    \centering
    \includegraphics[width=1.0\linewidth]{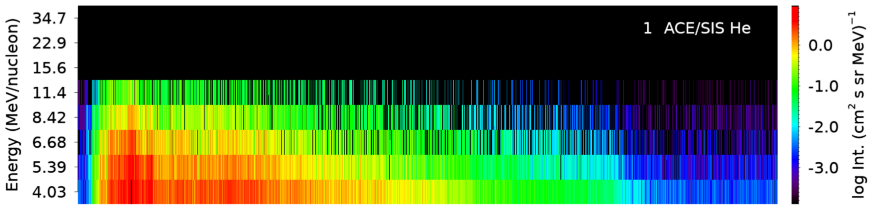}
    \includegraphics[width=1.0\linewidth]{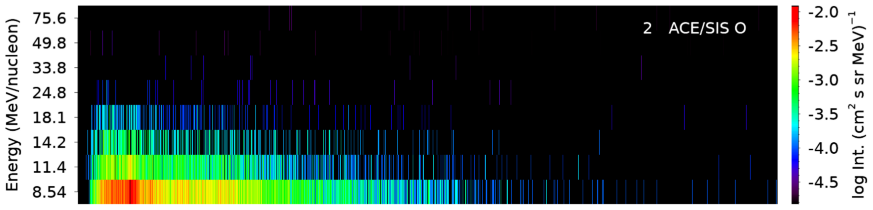}
    \includegraphics[width=1.0\linewidth]{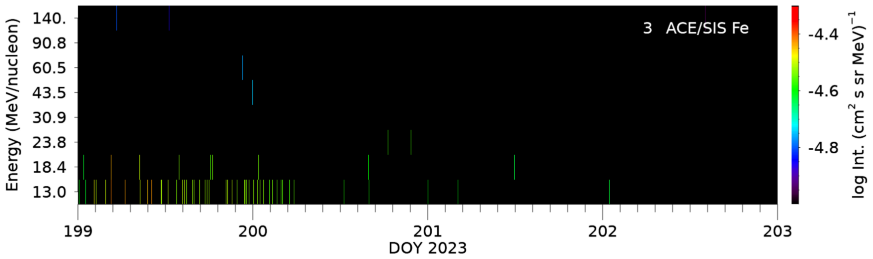}   
    \caption{Energy spectrograms measured by ACE/SIS for: (1) helium, (2) oxygen, and (3) iron.}
    \label{fig:ace_sis}
\end{figure}

Energy spectrogram plots, shown in Figures \ref{f-psp_let_spectra}, \ref{fig:stereo_let}, and \ref{fig:ace_sis}, display the flux at a given energy as a time profile throughout the entire SEP event. The fluence spectra are calculated for each spacecraft by integrating the energy spectrograms over the duration of the event: 17 July 2023 23:02 to 21 July 2023 23:02 for PSP and 18 July 2023 00:00 to 22 July 2023 00:00 for STA and ACE. In the case of PSP/LET, only the sunward-facing A side was included for fluence.  All three spacecraft observed a multi-day SEP event with a relatively rapid onset and an exponential decay after the peak intensity. The associated CME-driven shock crossed PSP at 19 July 2023 04:59 and later reached STA \& ACE at 20 July 2023 15:29, but there was no coincident increase in particles intensities at any of the spacecraft in this study's measured energy range. This suggests the shock had weakened and was no longer accelerating significant numbers of MeV-energy ions at/beyond 0.65 au.

\begin{figure}[htbp]
    \centering
    \includegraphics[width=1.0\linewidth]{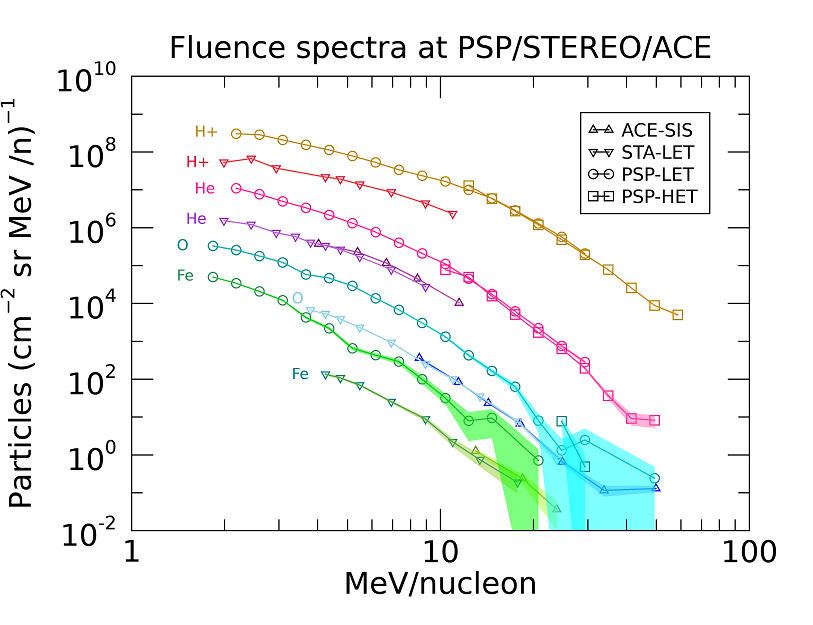}
    \caption{Comparison of fluence spectra for protons, helium, oxygen, and iron ions obtained from PSP/HET, PSP/LET, ACE/SIS, and STA/LET instruments during the SEP event.  Shading indicates the statistical uncertainties.}
    \label{fig:fluence_comparison}
\end{figure}

Figure \ref{fig:fluence_comparison} illustrates the 4 day integrated fluence measurements for protons, helium, oxygen, and iron ions obtained from \isois , ACE/SIS, and STA/LET instruments (spectrograms for the same time period are shown in Figures \ref{f-psp_let_spectra}, \ref{fig:stereo_let}, \ref{fig:ace_sis}).  The agreement between STEREO and ACE for He, O and Fe is good.  It is clear that the spectra are not simple power-laws, but gradually steepen with increasing energy.  Despite the separation between PSP, STEREO, and ACE, the spectra are remarkably similar in shape, suggesting little impact on the shape from transport.

To compare fluence measurements across instruments while accounting for differences in spacecraft distances, we initially applied radial adjustment factors to the fluence spectra for each ion species to account for PSP being located at 0.65 au from the Sun, while STEREO and ACE were positioned at 0.96 and 1.01 au near Earth, respectively. We scaled the fluence spectra for each ion species in a manner similar to \cite{lario2006} for the $r$-dependence based on Helios and IMP-8. Initially, an inconsistency in across different ion species was noticed when scaling the fluence of heavy vs. light ions for single $r^n$ value. Instead, we found it necessary to use different $r^n$ values for light and heavy ions.

\begin{figure}[htbp]
    \centering
    \includegraphics[width=1.0\linewidth]{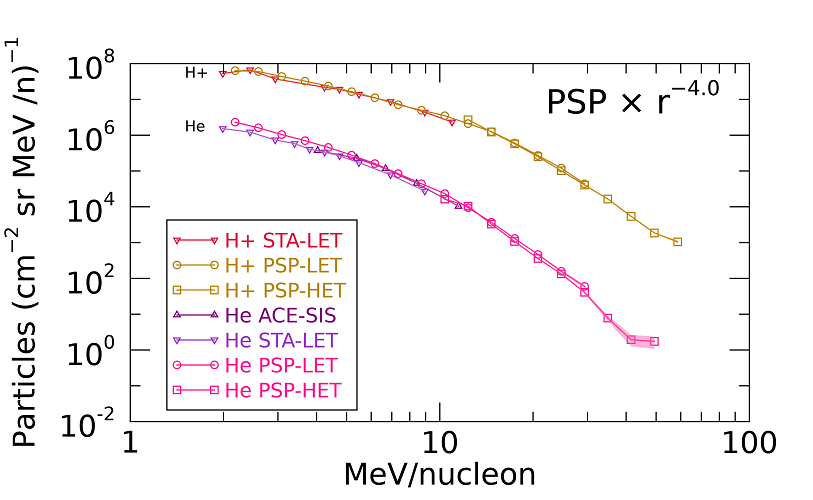}
    \caption{Fluence spectra for hydrogen and helium ions from the three spacecraft, scaled by a factor of $r^{-4}$ .}
    \label{fig:radial_adjustment_light}
\end{figure}

\begin{figure}[htbp]
    \centering
    \includegraphics[width=1.0\linewidth]{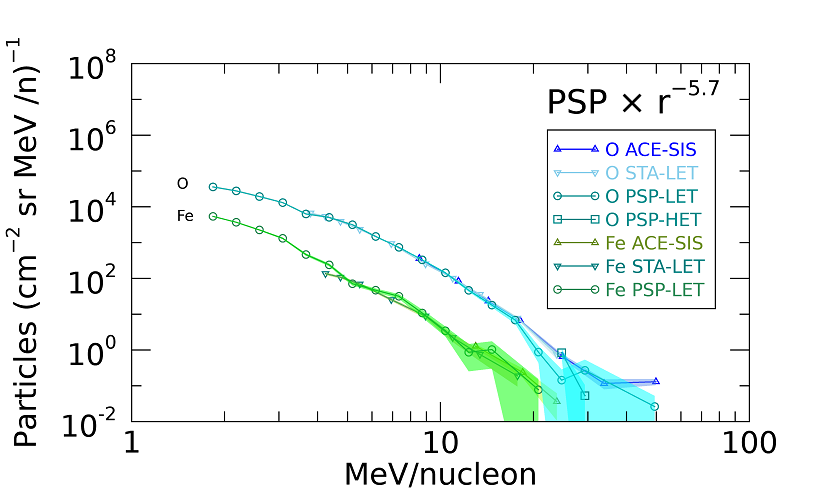}
    \caption{Fluence spectra for oxygen and iron ions from the three spacecraft, scaled by a factor of $r^{-5.7}$ .}
    \label{fig:radial_adjustment_heavy}
\end{figure}

Figure \ref{fig:radial_adjustment_light} illustrates the radial adjustments applied to the fluence spectra for hydrogen and helium ions, the PSP/HET and PSP/LET fluence spectra are scaled by a factor of 4.76, corresponding to $r^{-4.0}$, where $r$ represents the radial distance from the Sun to each spacecraft.  Figure \ref{fig:radial_adjustment_heavy} depicts the radial adjustments applied to the PSP/HET and PSP/LET fluence spectra for oxygen and iron ions. These heavier ions exhibit a stronger depletion in fluence abundance at the near-Earth spacecraft, requiring a radial adjustment of a factor of 9.23, corresponding to $r^{-5}$ .

\subsection{Ion Composition} \label{ion_comp_diff}
\begin{figure}[htbp]
    \centering
    \begin{overpic}[width=1.0\linewidth]{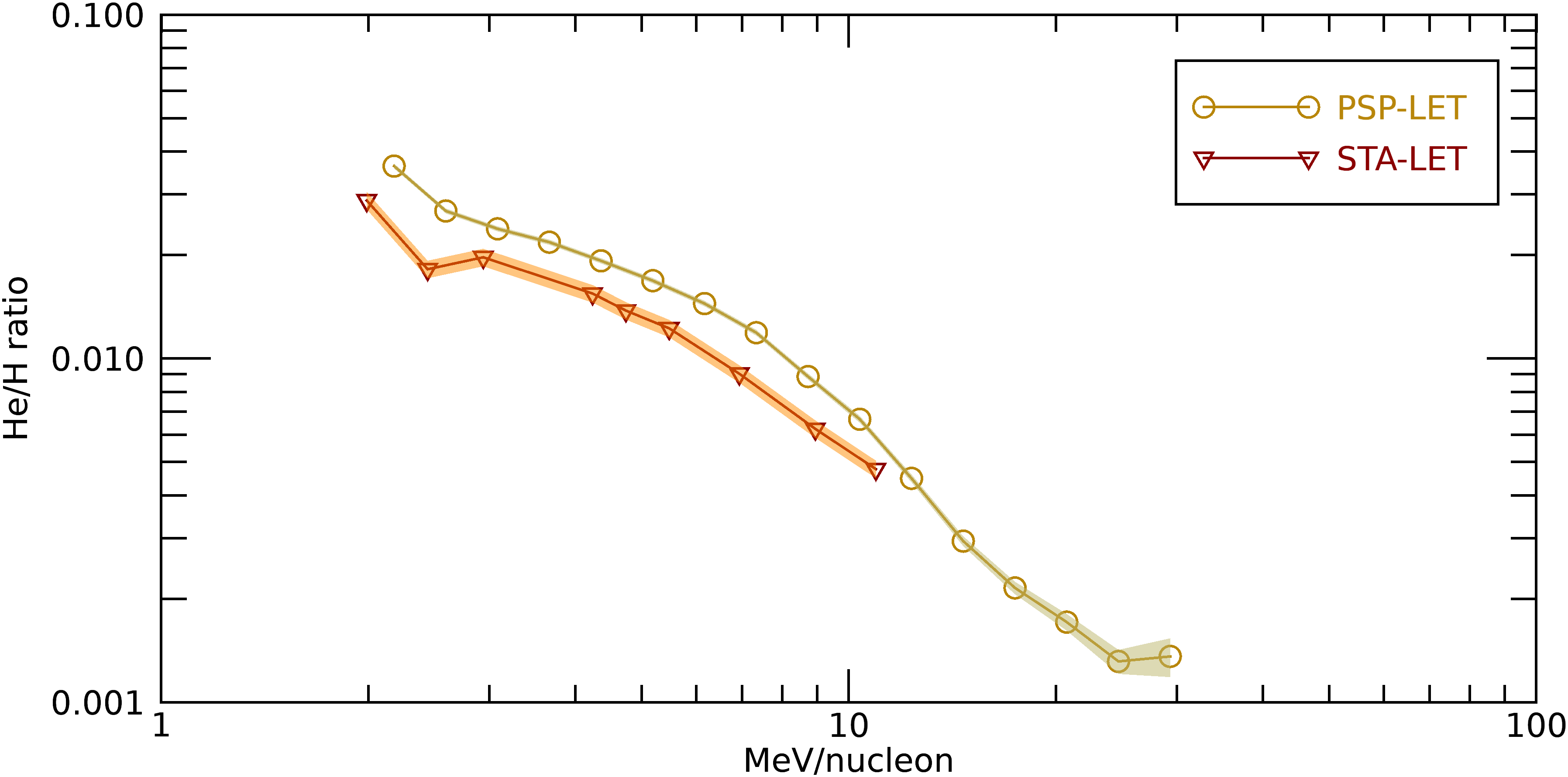}
        \put(15,42){\color{black}\large A}
    \end{overpic}
    \begin{overpic}[width=1.0\linewidth]{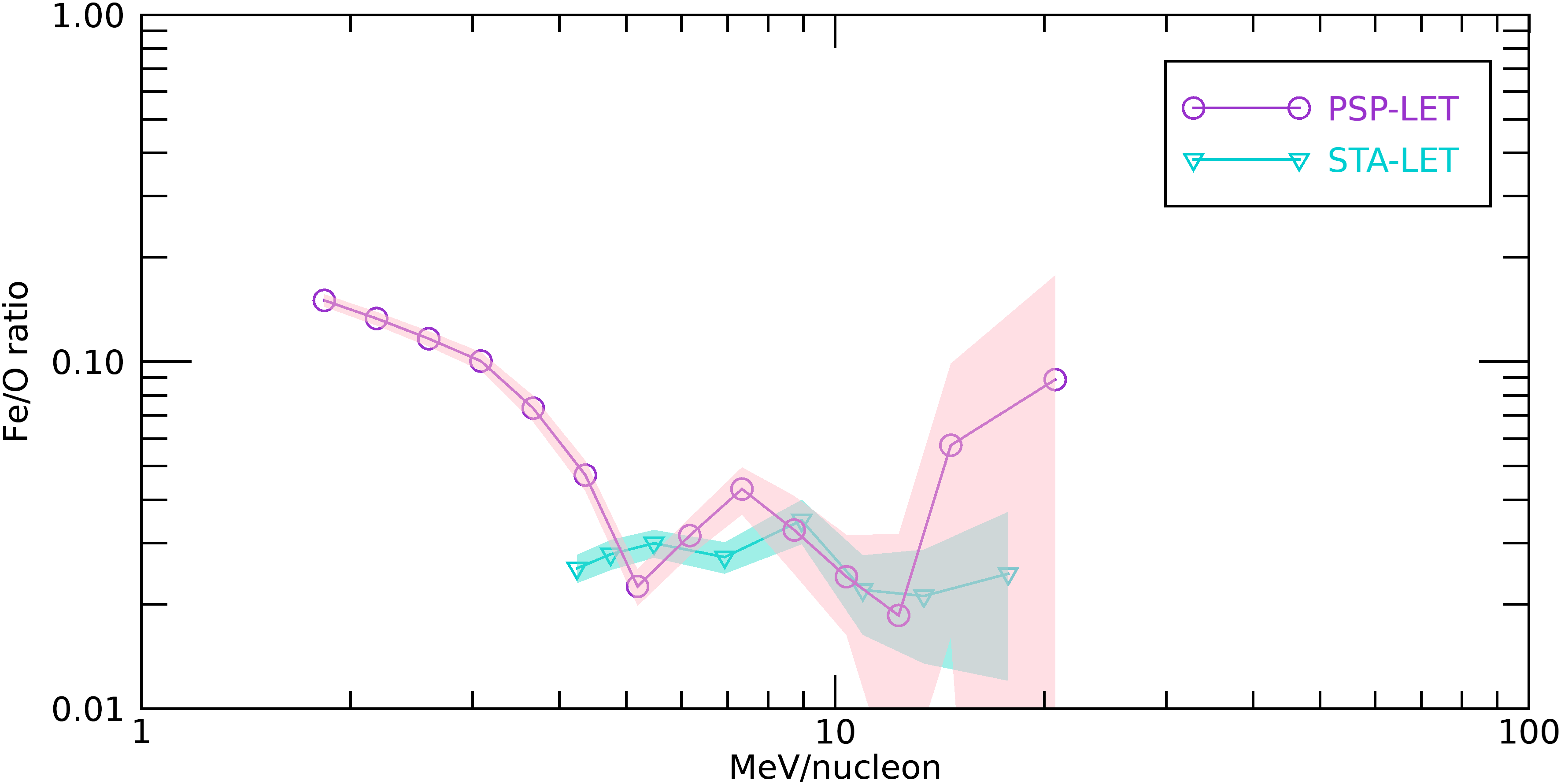}
        \put(15,42){\color{black}\large B}
    \end{overpic}
    \caption{Abundance ratios of (A) He/H and (B) Fe/O for selected energies integrated over the SEP event.}
    \label{fig:ion_ratio}
\end{figure}

Ion composition ratios, such as the ratio of helium to hydrogen (He/H) and the ratio of iron to oxygen (Fe/O), quantify the relative abundances of ion species in the overall energetic particle population and their evolution throughout SEP events. Figure \ref{fig:ion_ratio} depicts He/H and Fe/O for overlapping energies from STA/LET and PSP/LET A.  At the lowest energies (around 2 MeV/nucleon) both the He/H and Fe/O abundance ratios are fairly typical of large SEP events \citep{reames1995} and then decrease with increasing energy.  Such behavior is often observed in SEP events as a result of species-dependent breaks in the fluence spectra \citep{cohen2005,mewaldt2005}.  The He/H abundances are slightly lower at STEREO compared to PSP, but the energy dependence is very similar.  Although only the PSP Fe/O measurements extend to lower energies, where they overlap with the STEREO measurements the ratios are approximately the same and relatively energy-independent.  It is only the lower energies of the PSP observations that exhibit a strong energy dependence.  

\section{Discussion}

The radial gradients measured in this SEP event exceed that of prior studies by \cite{lario2006}, which estimated protons to have a radial dependence of $r^{-3.3}$ over similar heliospheric distances. For the 17 July 2023 event, we found proton and helium radial dependence to be near $r^{-4}$ which suggests more scattering has occurred throughout transport and is more similar to protons modeled by \cite{kozarev2010} at $r^{-3.8 \pm{0.3} }$, albeit at higher energies of $>$80 MeV. One source of increased scattering could be interactions of CMEs as they merge together.  As mentioned earlier a small, narrow CME preceded the halo CME imaged in Figure \ref{f-x-ray_cme}.

\begin{figure}[htbp]
    \centering
    \includegraphics[width=1.0\linewidth]{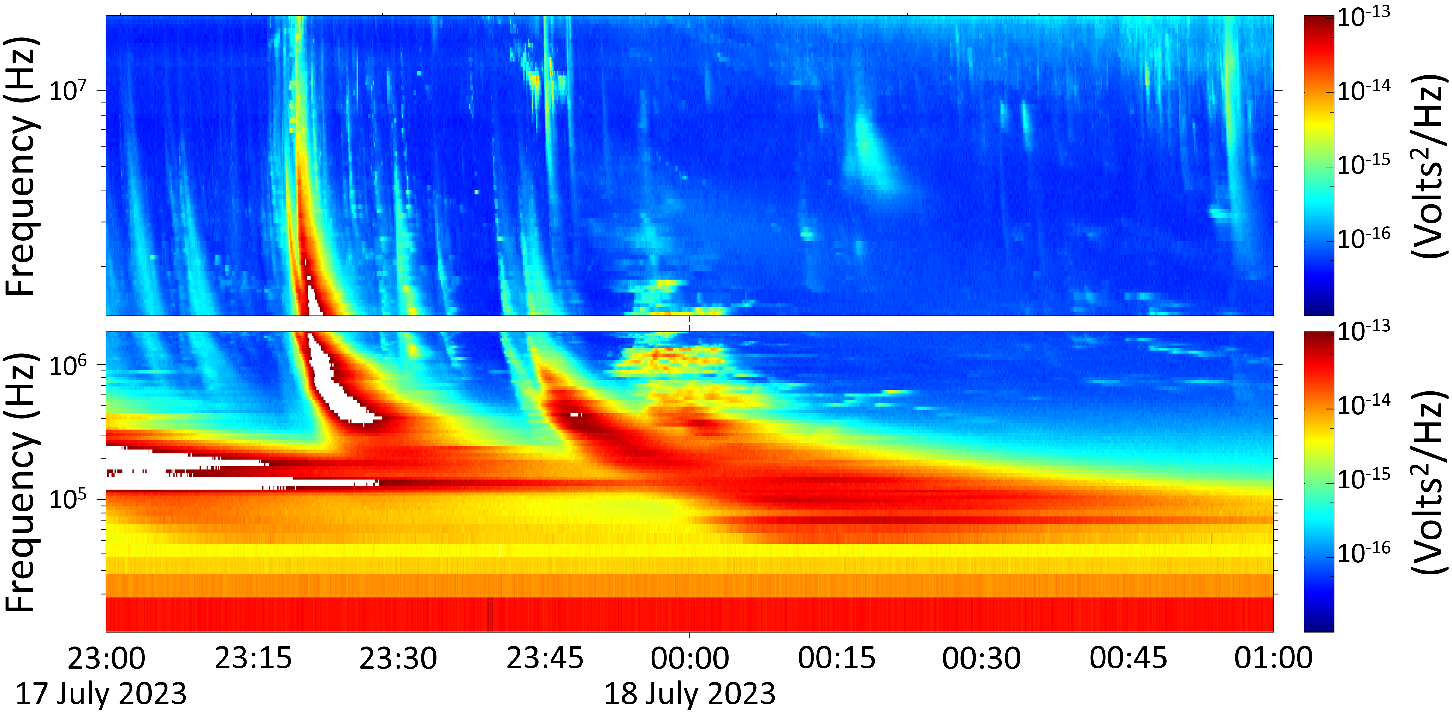}
    \caption{Time-series of radio band intensities, from the FIELDS instrument onboard PSP, showing Type II, III, and IV radio bursts occurring at of near solar activity.}
    \label{fig:radio_data}
\end{figure}

\begin{figure}[htbp]
    \centering
    \begin{interactive}{animation}{movie_CHASE_SUVI.v2}
    
    \begin{overpic}[width=0.27\linewidth]{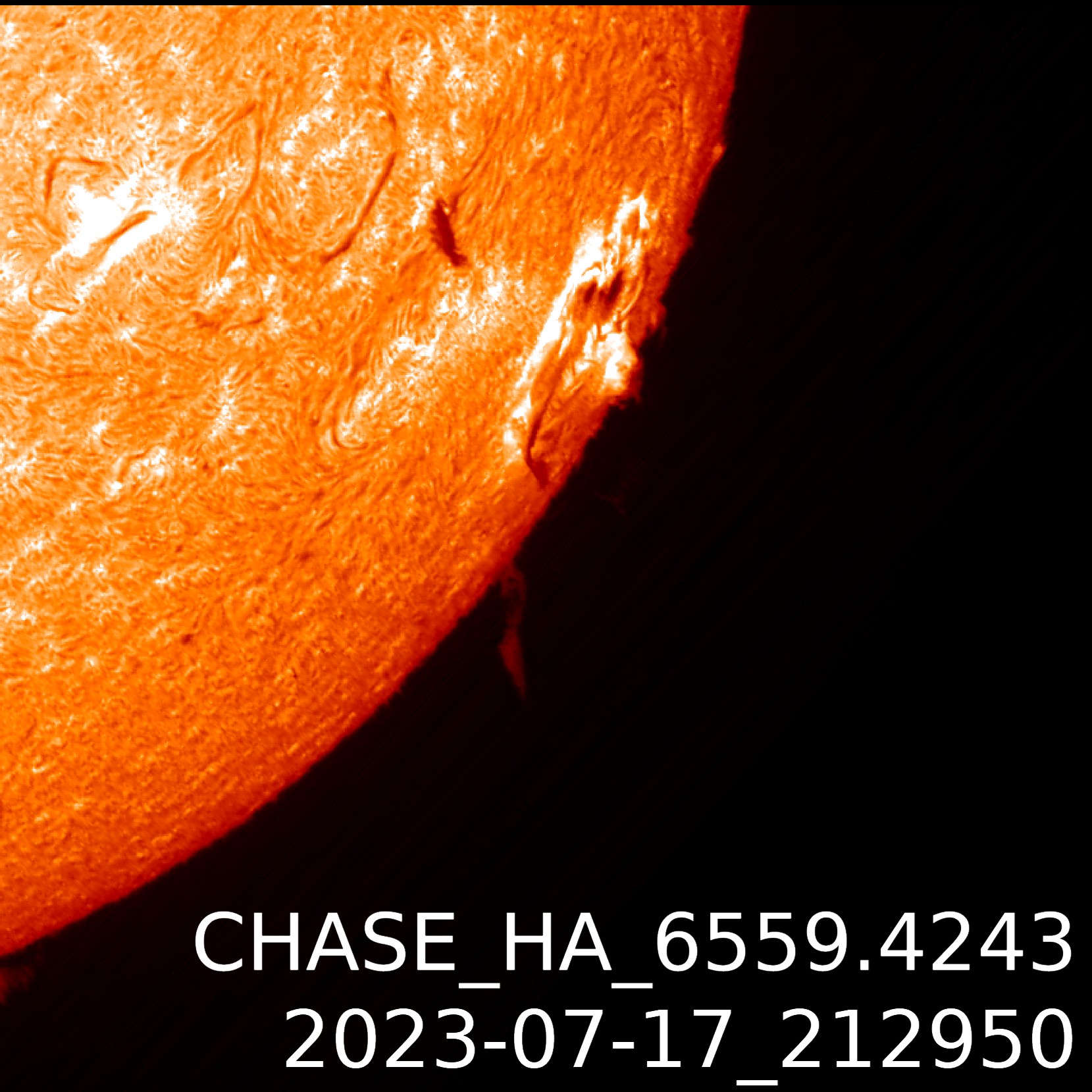}
        \put(85,85){\color{white}\large 1}
    \end{overpic}
    \begin{overpic}[width=0.27\linewidth]{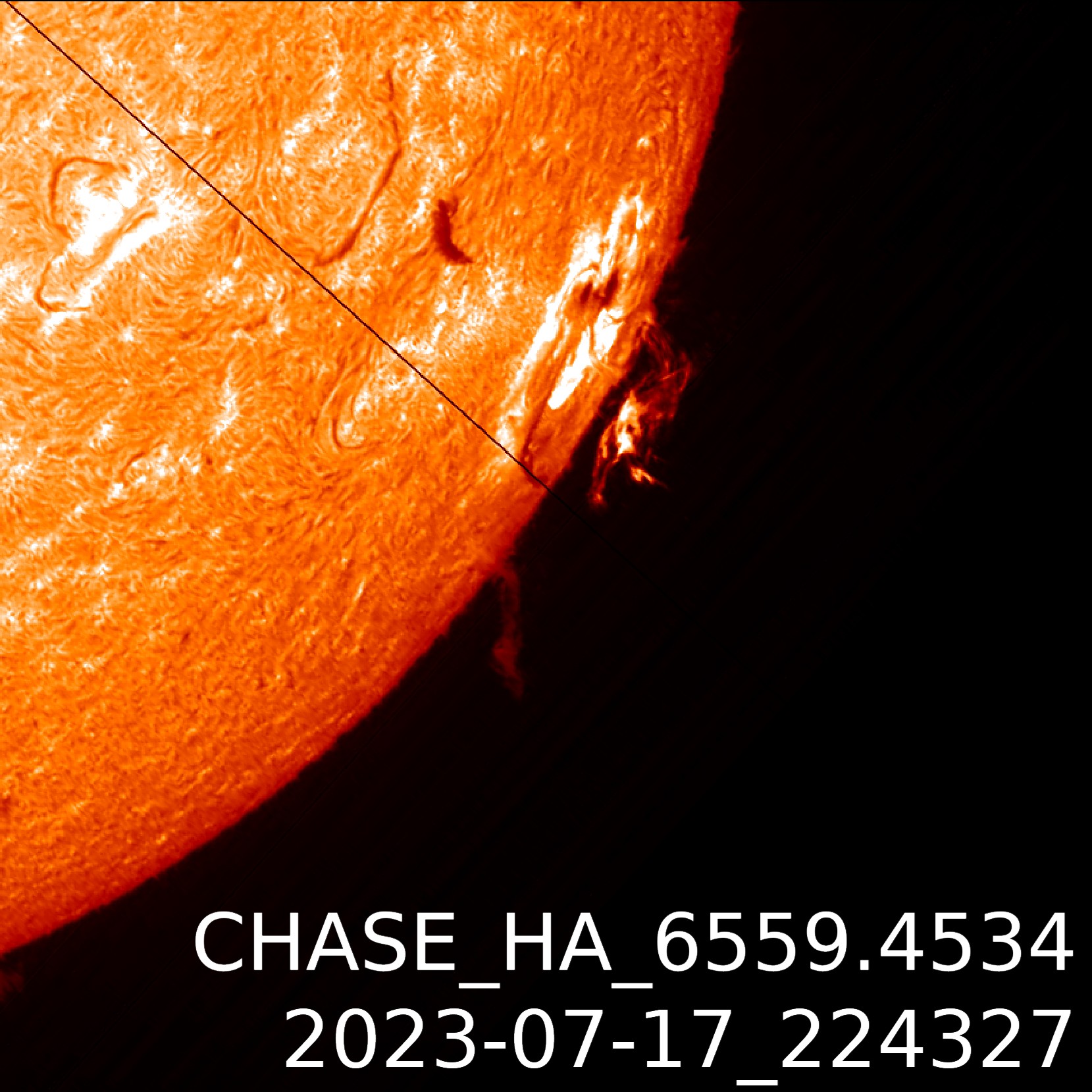}
        \put(85,85){\color{white}\large 2}
    \end{overpic}
    \begin{overpic}[width=0.27\linewidth]{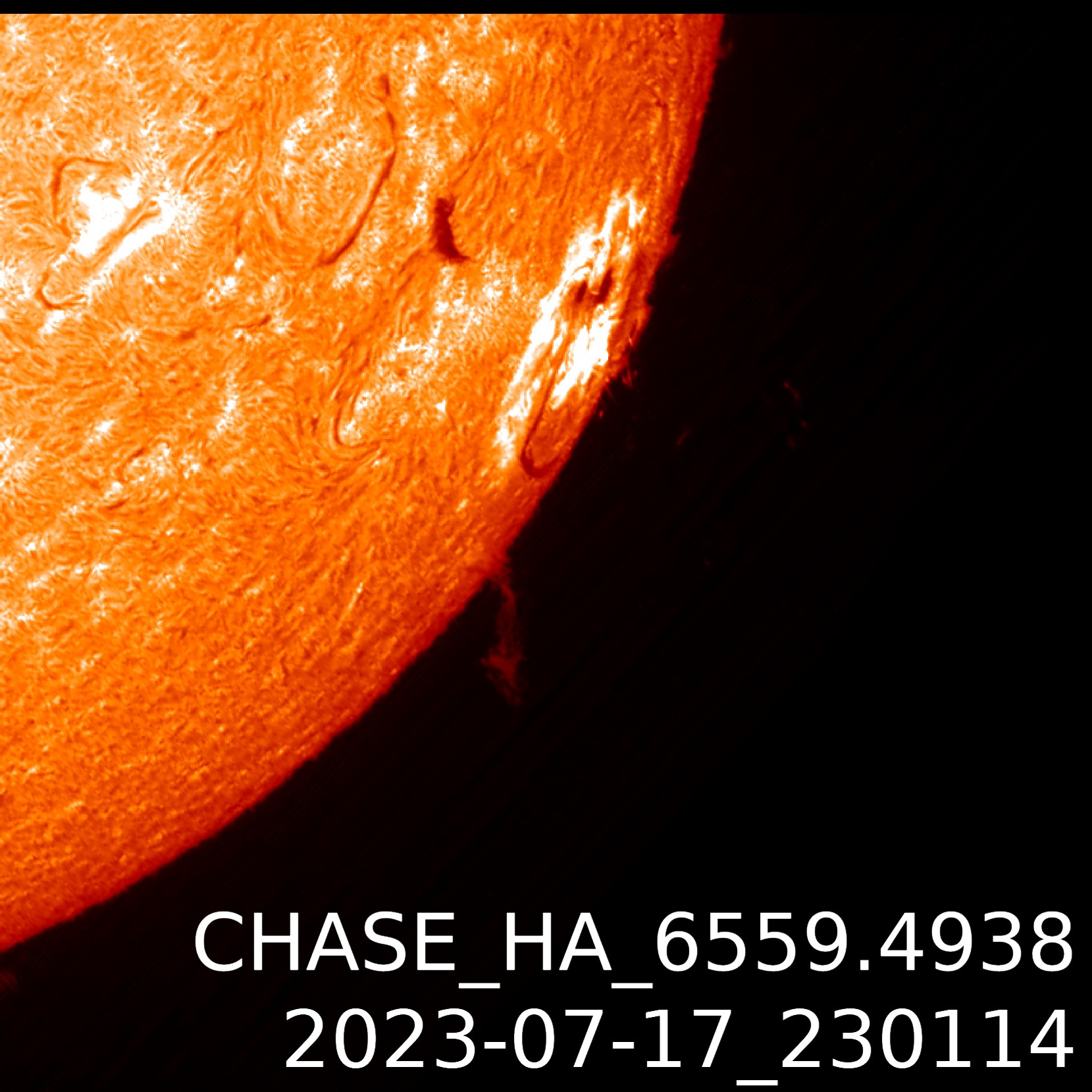}
        \put(85,85){\color{white}\large 3}
    \end{overpic} \\
    \begin{overpic}[width=0.27\linewidth]{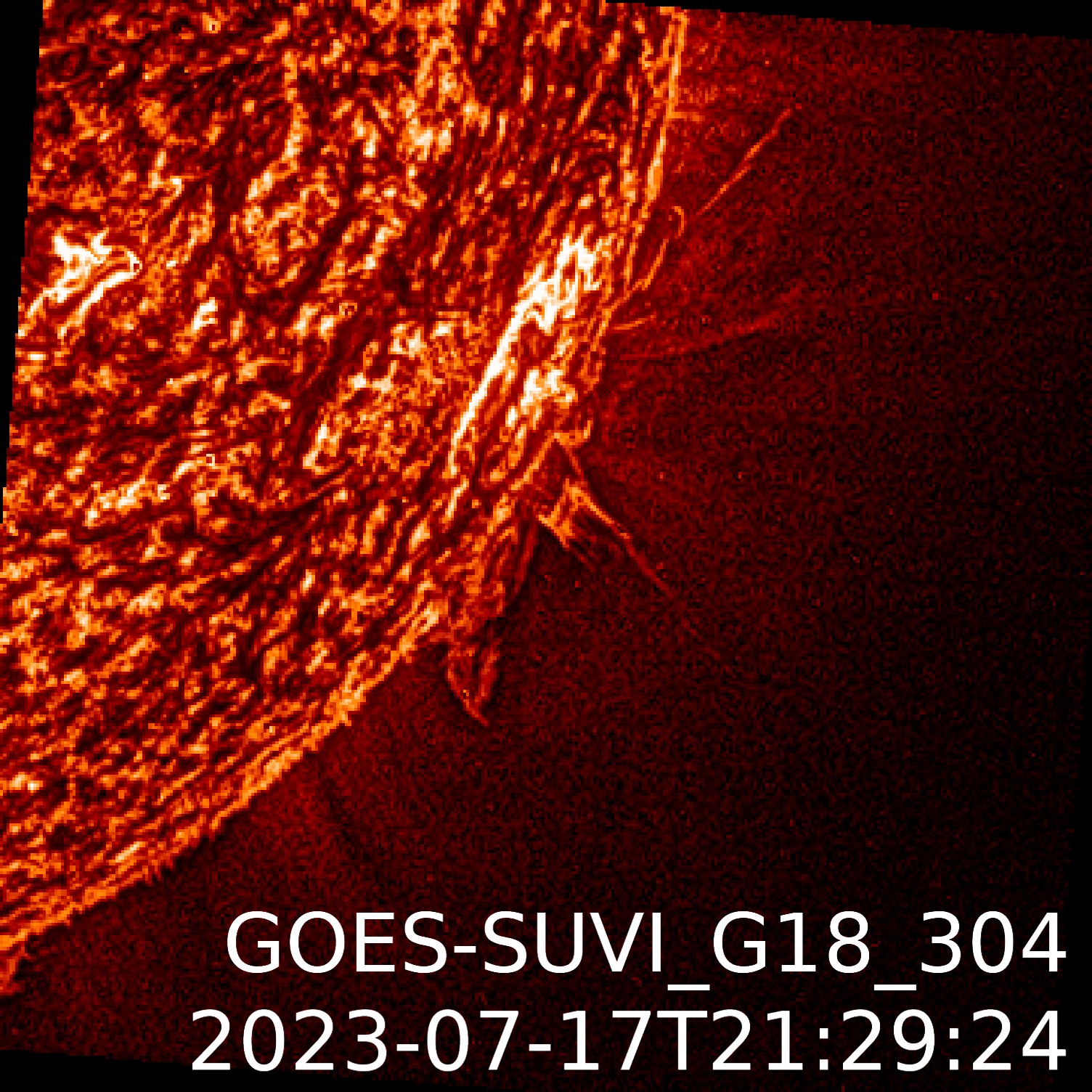}
        \put(85,85){\color{white}\large 4}
    \end{overpic}
    \begin{overpic}[width=0.27\linewidth]{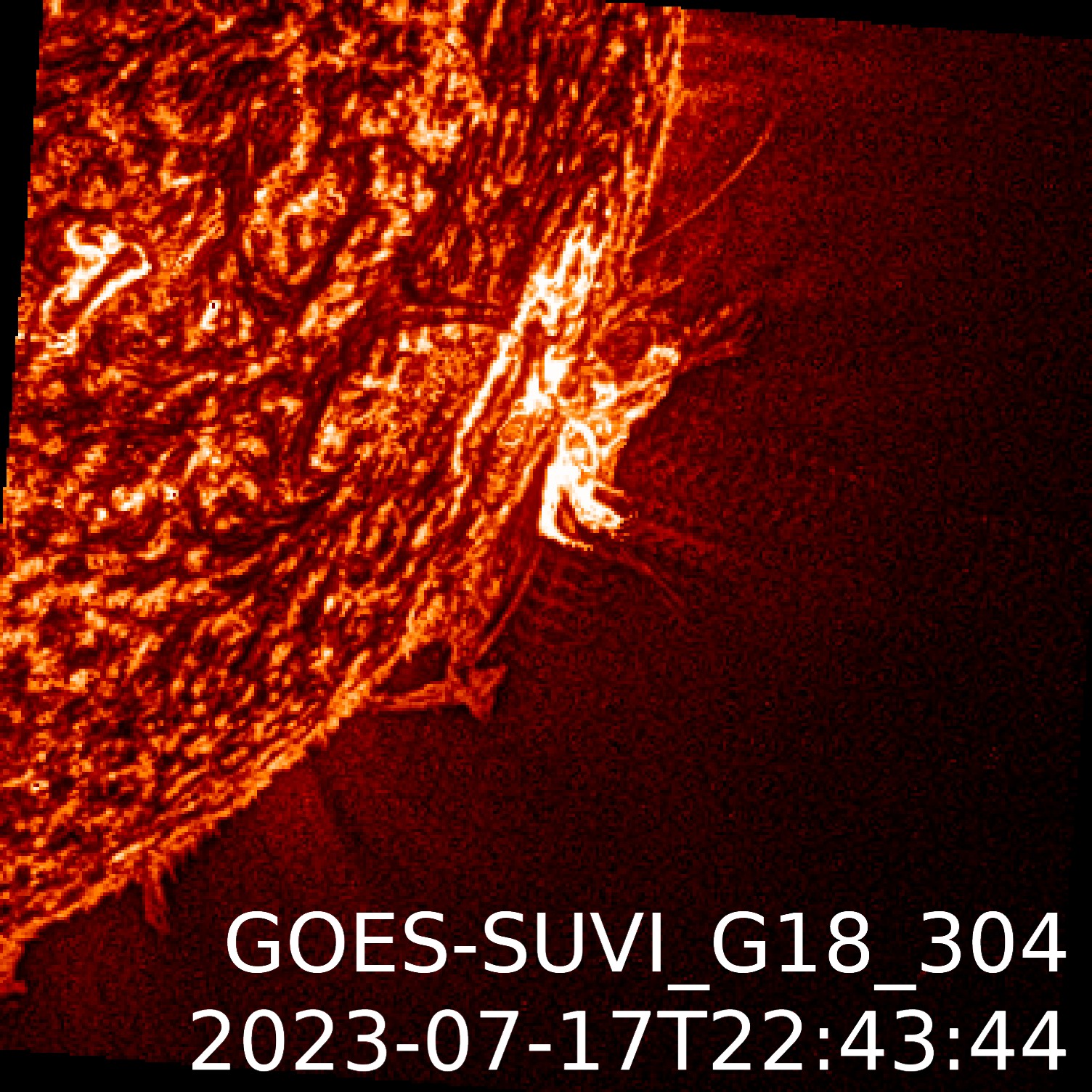}
        \put(85,85){\color{white}\large 5}
    \end{overpic}
    \begin{overpic}[width=0.27\linewidth]{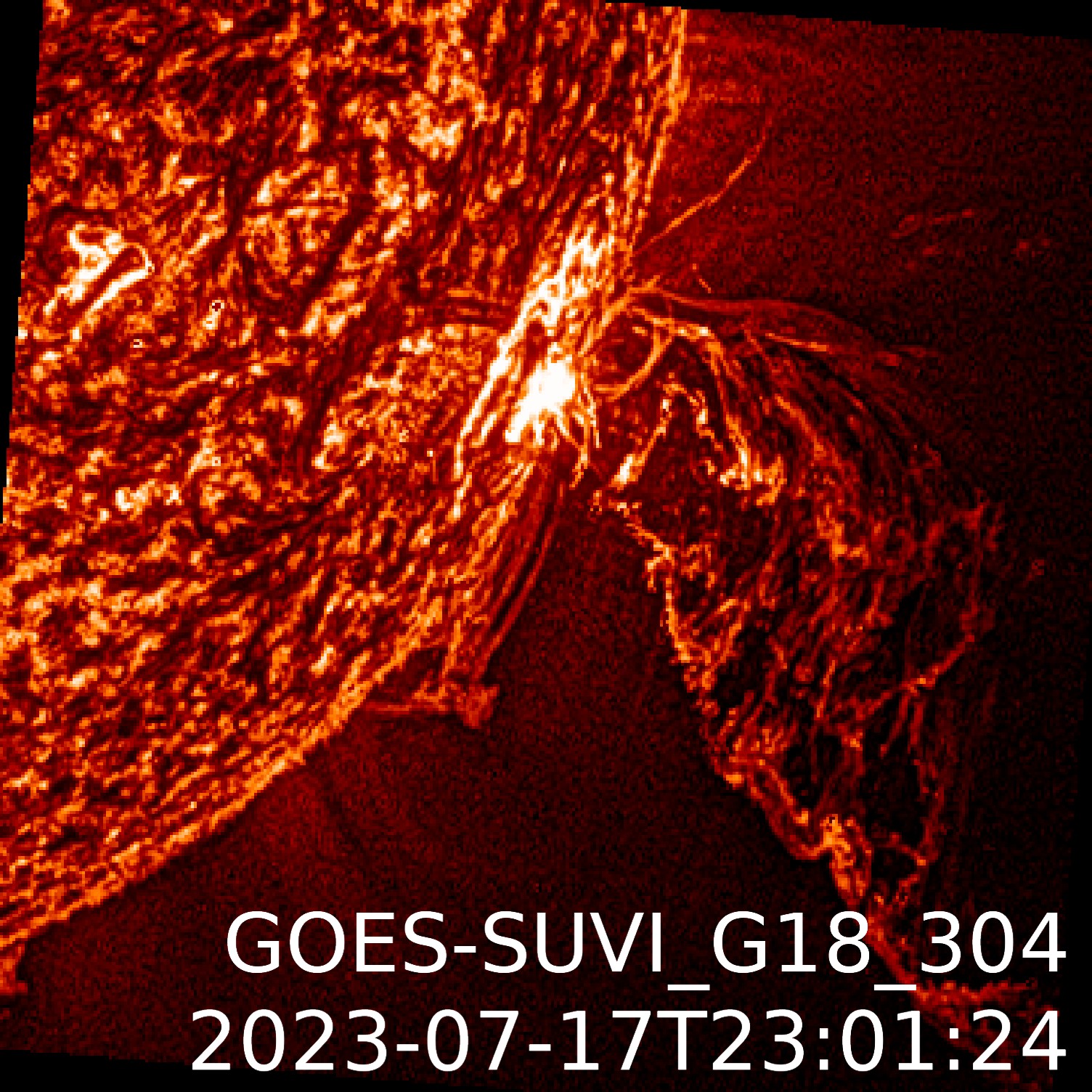}
        \put(85,85){\color{white}\large 6}
    \end{overpic}
    \end{interactive}
    \caption{Solar atmosphere images in H-alpha from the CHASE spacecraft show the progression (1) prior to first flare, (2) during the initial flare, where a prominence formed above AR 13363, and (3) the prominence being ejected. The atmosphere is also imaged in 30.4 nm by GOES/SUVI 18, which shows the atmosphere (4) prior to first flare, (5) during the initial flare, where a prominence formed above AR 13363, and (6) the prominence being ejected. Note: The supplemental animation begins on 17 July 2023 at 19:26 and continues over the next 4 hours to show a prominence quickly forming and being ejected outward from AR 13363. The GOES/SUVI 18 animation is continuous, but the CHASE animation includes gaps and distortions due to data availability.}
    \label{f-imagers}
\end{figure}

Figure \ref{fig:radio_data} shows radio observations over the 120 minute period centered around the CME-CME interaction time. A Type III solar radio burst is clearly seen taking place nearly coincident with the M5.6 flare that peaked at 23:17, noted by the fast drifting signal from high to low frequencies over the course of the next 15 minutes. Type III solar radio bursts are associated with the strong flares accelerating electrons which stimulate radio emission. From 23:41 to 23:47, a Type IV radio burst at 10 MHz occurs during the period of initial prominence eruption shown in Figure \ref{f-imagers}. Type IV radio bursts have been previously understood to be narrow beams emitting from coronal loops overlying the EUV post-eruption arcades and associated with CME and prominence eruptions \citep{gopalswamy2016}. Faintly drifting downward from 23:51 through the end of Figure \ref{fig:radio_data} at 01:00 is a Type II radio burst, which is associated with the leading edge of a CME due to shock waves accelerating electrons to stimulate radio emission. Curiously, at 23:51 there is a 14-minute \app1 MHz enhancement which is coincident with the timing of the large, fast halo CME, which emerged from AR 13363 in Figure \ref{f-x-ray_cme}, overtaking the initial narrow, slower CME.

Interestingly, the oxygen and iron radial dependence was considerably stronger at $r^{-5.7}$ compared to lighter ions. The cause of this is not immediately clear, although one distinct difference between the heavy and light ions is their typical charge-to-mass (Q/M) ratios. In particular H and He are expected to be fully stripped for a range of coronal temperatures while O and Fe are not.  This could lead to a significant difference in the relative rigidities and therefore the transport of the heavy ions relative to the lighter ions from PSP to STEREO \& ACE.  Potentially relevant to this is the appearance of prominence material near AR 13363 between the eruption of the two M-class flares.

Figure \ref{f-imagers} provides low coronal context of the magnetic evolution occurring in AR 13363 via H-alpha images from CHASE \citep{li2022} and ultraviolet 30.4 nm images from GOES/SUVI 18 \citep{darnel2022}, which correspond to emission of H at \app5000 K and He II at \app50000 K, respectively. This magnetic evolution shows the moments before the first M5.0 flare, the formation of a prominence and its ejection during the beginning at the beginning moments of the M5.6 flare. The M5.0 flare coincided with a relatively small, narrow CME while the second flare triggered a massive halo CME that traveled at a significantly higher speed and later dominated the COR2 field of view, shown in Figure \ref{f-x-ray_cme}. 

Of particular interest is the movement of an over-the-limb prominence, at the lower right of the solar disk, to a location just above the M5.6 flare moments before its eruption. The series of images in Figure \ref{f-imagers} shows that prior to the first M5.0 flare, no cool prominence material was located above AR 13363. Then at the time of the M5.0 flare, cool prominence material has moved to the region for a short period of time before it is violently ejected away. The location of this prominence raises the possibility that it was injected into the seed particle source for the subsequent SEP event. Soft x-ray flare emission is highly efficient at ionization of particles within the flare reconnection loops, which we regularly see as emission from ionized solar plasma in extreme ultraviolet images. Thus, the cooler prominence, of which a typical temperature is between 0.005 to 0.05 MK, at heights above the flare loops acting as a seed particle source would inject mostly O and Fe ions with considerably lower charge-to-mass ratio  (Q/M) compared to other flare accelerated plasma. 

While intriguing, it is not clear that the lower rigidity (potentially exacerbated by the prominence material) of the heavy ions is the cause of the greater depletion in heavy ions along the final \app0.33 au of radial distance.  One would expect such a rigidity effect to also be reflected in a change in spectral shape from PSP to STEREO and ACE, with a greater depletion at the higher energies.  However, as can be seen from Figures \ref{fig:radial_adjustment_light} and \ref{fig:radial_adjustment_heavy}, common scaling does remarkably well for all of the energies measured by the three spacecraft.  It is unclear what mechanism could result in a radial gradient that is potentially mass or Q/M dependent (but not rigidity dependent) and is consistent with the results presented here. 

It is possible, however, that a change in spectral shape may be more evident at lower, or higher, energy ranges than covered within this study and this will be examined in future work. We have also explored a qualitatively similar multi-spacecraft SEP event between PSP, STEREO, and ACE on 7 August 2023 and did not see a similar radial dependency based on ion species, which may suggest that this is not a common observation and unlikely to be due to intercalibration issues.

Diffusive shock acceleration (DSA) tends to be more efficient at accelerating high Q/M ratio ions than those with low charge-per-mass according to \cite{lee1982,desai2003}. In the case of light ions like hydrogen and helium which are fully ionized, DSA can more easily elevate their energies to many MeV. However, heavy ions like oxygen and iron require significant heating to become fully ionized and thus are less efficiently accelerated by DSA. In the case of this particular SEP event, heavy ions with even lower ionization states from the cool prominence material may have been injected into the seed population and reduced the efficiency of DSA disproportionately as compared to H and He. 

A similar study of radial dependence of SEP events from observers with close footpoints, \cite{lario2006}, found that even though the magnetic connection between IMP 8 and Helios 1 was close, the steep radial dependence of peak intensities was enough to suggest that the spacecraft were not really magnetically connected. In this present study between PSP and STA, the release time and path length derived from VDA, as well as the derived magnetic footpoints of the spacecraft, suggest good magnetic connectivity between the spacecraft. The substantially different radial dependence of the heavy ions relative to the light ions suggests greater diffusion or scattering of heavy ions, but without a significant change in the qualitative spectral shape.

The curved, no-single power-law spectral shape might be explained by an efficient particle escape \citep{fraschetti2021}, but the steepening of the radial gradient for all ions and the surprising differing radial gradient for heavy vs. light ion fluence remains a challenge to explain. Modeling efforts can address the steep radial gradient by numerically solving the focused transport equation to determine the contribution of the focusing term to SEP radial gradient \citep{ruffolo1995,ruffolo1998}. Although, another approach which considers drift effects is ideal for modeling heavy ions separately. Drift velocities depend on the Q/M ratio, and SEP heavy ions are typically not fully stripped of electrons \citep{klecker2006}, so it is expected that heavy ions will have larger drift velocities than light ions at the same MeV/nucleon \citep{dalla2013}. Other considerations such as SEP pitch angle, magnetic field inhomogeneity, and turbulent scattering can further exacerbate the drift effects \citep{marsh2013,laitinen2023}.

\section{Conclusion}
The 17 July 2023 SEP event provided an ideal spacecraft alignment for observing how the SEP event characteristics change from 0.65 to \app1.0 au. The spectral shapes of each ion species observed at different spacecraft locations were qualitatively similar and a simple power-law in radial distance scaling between the PSP/HET, PSP/LET, STA/LET and ACE/SIS measurements worked remarkably well for the full energy range examined. While the H and He gradients were stronger than those found in previous studies using Helios and IMP-8, they are consistent with some modeling efforts with enhanced turbulence.  Such conditions may have resulted from a CME-CME interaction early in this event.  Quite surprisingly, the radial gradient of O and Fe was significantly stronger than that for H and He.  It is not clear what mechanism to attribute this to, but the similarity in spectral shapes observed by the different spacecraft suggests rigidity-related processes are not the primary cause.  The  abundance ratios of He/H and Fe/O were relatively unchanged from 0.65 to 1 au and exhibited the fairly common decrease with increasing energy, although this was limited to mostly below a few MeV/nucleon for Fe/O.

\section*{Acknowledgements}
We acknowledge the contributions of the Parker Solar Probe, STEREO, and ACE mission teams in collecting and providing data for this study. Parker Solar Probe was designed, built, and is now operated by the Johns Hopkins Applied Physics Laboratory as part of NASA's Living with a Star program, contract No. NNN06AA01C. Support from the LWS management and technical team has played a critical role in the success of the Parker Solar Probe mission. We thank the scientists and engineers whose technical contributions prelaunch have made the \isois\ instruments such a success. A. Vourlidas was supported by NASA grant No. 80NSSC22K0970. As always, we thank the Sun for providing the energy and particles necessary for this inquiry.


\end{sloppypar}

\bibliography{refs.bib}
\bibliographystyle{aasjournal}



\end{document}